\documentclass[twocolumn,showpacs,preprintnumbers,amsmath,amssymb]{revtex4}
\usepackage{graphicx}
\usepackage{bm}
\usepackage{setspace}
\def\tagform@#1{\maketag@@@{\normalfont\bf\normalcolor[\ignorespaces#1\unskip\@@italiccorr]}}

\def\be{\begin{equation}}
\def\ba{\begin{eqnarray}}
\def\a{\alpha}
\def\b{\beta}
\def\g{\gamma}

\def\d{\delta}

\def\h{\eta}

\def\m{\mu}

\def\x{\xi}
\def\p{\pi}

\def\f{\varphi}

\def\Ps{\Psi}

\def\bF{{\mathbf F}}

\def\bx{{\mathbf x}}
\def\bX{{\mathbf X}}
\def\bv{{\mathbf v}}

\def\br{{\mathbf r}}

\def\bxi{\mbox{\boldmath{$\x$}}}

\def\ee#1{\label{#1}\end{equation}}
\def\ea#1{\label{#1}\end{eqnarray}}

\begin{document}
\title{Chiral separation in microflows}
\author{ Marcin Kostur, Michael Schindler, Peter Talkner, Peter H\"anggi}
\affiliation{Universit\"at Augsburg, Institut f\"ur Physik, 
Universit\"atsstrasse 1, D-86135 Augsburg, Germany}
\date{\today}
\begin{abstract}
Molecules that only differ by their chirality, so called enantiomers,
often possess different properties with respect to their biological
  function. Therefore, the separation of enantiomers presents a prominent
  challenge in molecular biology and belongs to the ``Holy Grail'' of
  organic chemistry. We suggest a new separation technique for
  chiral molecules that
  is based on the transport properties in a microfluidic flow with spatially
  variable vorticity. Because of their size the thermal fluctuating
  motion of the molecules must be taken into account.  These
  fluctuations play a decisive role in the proposed separation mechanism.
\end{abstract}
\maketitle
The main established methods of enantiomer separation are gas or
liquid chromatography and capillary electrophoresis
\cite{Ahuja}.
In both cases specific chiral 
filling materials are used to obtain different elusion times of the
enantiomers. Moreover, long columns are needed through which the
enantiomers 
are dragged by a large pressure or a high voltage
difference. Chemically less specific methods that work with less
substance and that  
do not require high voltages or pressure would be of
great advantage.

During the last ten years the development of microfabricated fluid
devices 
has experienced 
enormous progress \cite{Figeys}. 
New ideas how
to manipulate small amounts of fluids and substances therein have been
suggested \cite{Bruin}. 
Recently, it has been demonstrated that
the agitation of fluids by surface acoustic waves on
piezoelectric substrates presents a versatile method of manipulating
and controlling the flow of small amounts of a fluid 
\cite{Guttenberg04,Guttenberg05}.

Various separation mechanisms have been proposed.
The separation according to size was
predicted for particles in a fluid that is periodically pumped
through an array of parallel 
cylindrical pores with 
ratchet shaped
cylinder radii \cite{Kettner,Astumian} 
and experimentally
corroborated in Ref.~\cite{Matthias}. 
de Gennes showed that a chiral crystal floating on a liquid
surface or gliding on a solid surface under the influence of an external force
will generally move in a direction that is specific for the chirality of the
crystal \cite{deGennes}.  
For molecules, de Gennes
argued thermal fluctuations would destroy this effect
\cite{deGennes}. In the present letter we propose a different mechanism that
uses the constructive influence of 
thermal noise in combination with the nonlinear dynamics of
molecules in a flow field to attain a separation
of opposite chiral partners.   
  
Inertial forces acting on a suspended point-like particle are
generally 
small \cite{Purcell} and can be neglected. Consequently, the 
particle is advectively
transported
with the 
fluid velocity
at the particle's actual position.    
For an incompressible fluid, the particle motion then is
volume conserving, and,
as a consequence,  
attractors do not exist \cite{Lichtenberg}.
For small particles, diffusion provides another transport mechanism,
which tends to level out concentration differences of suspended
particles. 
In the case of a point particle, diffusion in any incompressible flow
results in a homogeneous distribution if external forces like gravity
or electric fields are absent.

A contrasting 
picture results for extended particles. 
The local velocity of a surface point of an extended
particle need not coincide
with the fluid velocity that one would observe at this point in the
absence of the particle. As a consequence, the volume of the state
space spanned by the particle's degrees of freedom 
is no longer conserved by the dynamics, which,
as a result may display attractors for stationary flow fields, 
in spite of the fact that the dynamics is reversible \cite{Politi}. 
In general, several attractors will coexist. Which of them is
approached after sufficiently long time depends on the  initial
conditions.
If the particle is still small, say with a diameter of 1 $\m$m or
less,  
thermal noise at ambient temperature 
will destroy very weak 
attractors and populate
stronger ones, with weights depending on the stability of the
attractors.

In order to illustrate the working principle, we study a model
describing the motion of an extended {\it planar} object in a {\it two
dimensional} incompressible 
stationary flow. 
This object, or  
``molecule'' as we will call it, 
differs from its chiral partners only in the sequence of three
spherical ``atoms'' with different friction coefficients.
For a chiral molecule the mirror image does not match with the 
original
molecule upon any motion in the plane.
It is the dependence of the transport properties on the chirality
which is in the focus 
of this paper. 

At short times, the differences between the dynamics of chiral partners
possibly are  
rather small but  
at long times they will
lead to different attractors with different stability properties. 
With the omnipresence of thermal noise in the fluid 
these different stability
patterns will cause spatial distributions that are different for
chiral partners to an extent that 
they can be used for an effective chiral separation. This is a
robust mechanism that also works in three dimensions.

{\it Motion of an extended molecule in a flowing fluid.}
We consider three atoms with positions $\bx_i$, $i=1,2,3$ 
that are  
rigidly connected, 
see Fig.~1. 
\begin{figure}
\includegraphics{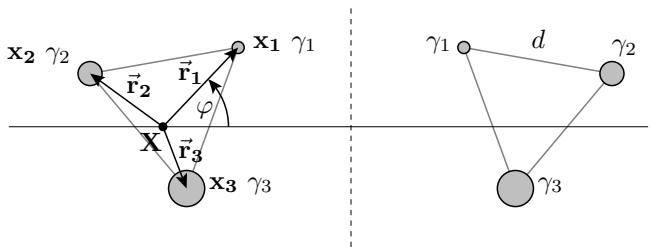}
\caption{A chiral molecule  
consisting of three atoms with different friction coefficients $\g_i$ 
on an equilateral triangle at positions $\bx_i = \bX
  + \br_i(\f)$. 
  The mirror particle differs only in
  the {\it sequence} of the friction coefficients. The rigid rods
  (gray lines) of lengths $d$ maintain the structure.  
}
\label{f1}
\end{figure}
The 
force on the
$i$th particle exerted by the fluid
moving with velocity $\bv(\bx)$ is assumed to be
proportional to the relative velocity of the particle $\dot{\bx}_i(t)$ 
with respect to the fluid, i.e., 
$
\bF_i^{\text{fl}}=\g_i \left ( \bv(\bx_i) -\dot{\bx}_i \right )
$  
where $\g_i$ denotes the friction coefficient of the $i$th atom, and
$\bv(\bx)$ is the velocity field of the fluid in absence of the atoms. 
In a fluid at temperature
$T$, the frictional forces are always accompanied by random
forces, which, for the sake of simplicity, are assumed to be  independent
Gaussian thermal white
noises, $\bxi_i(t)$, with
cartesian components $\x_{i,\a}(t)$, $\a=1,2$,
satisfying the fluctuation-dissipation relation   
$ \langle \x_{i,\a}(t) \x_{j,\b}(s)
\rangle = 2 \g_i \:k_B T\: \d_{i,j}\:\d_{\a,\b}\: \d(t-s)$.
We also have to take into
account forces $\bF_{i,j}$ caused by the $j$th atom and acting on the
$i$th one that maintain the rigid structure of the molecule. 
We neglect
hydrodynamic interactions between the individual atoms \cite{Dhont}
as well as small and very short-lived inertial terms \cite{Purcell}. 
The particle motion then follows from the balance of forces on each particle
\be
 \g_i \left [  \bv(\bx_i) -\dot{\bx}_i \right ] +
\sum_{j \neq i}\bF_{i,j} + \bxi_i(t)  = 0
\ee{dA}
The actual degrees of freedom  
are translations of the molecule as a whole,
and rigid
rotations. 
The momentary position 
of the molecule is conveniently specified by   
the {\it center of friction} $\bX = \sum_i \tilde{\g}_i \bx_i$ 
where $\tilde{\g}_i=  \g_i/\sum_j \g_j$ are normalized friction
coefficients, 
and by an angle $\f$ fixing the orientation, see Fig.~\ref{f1}.
\begin{figure}[!t]
\includegraphics[height=5cm]{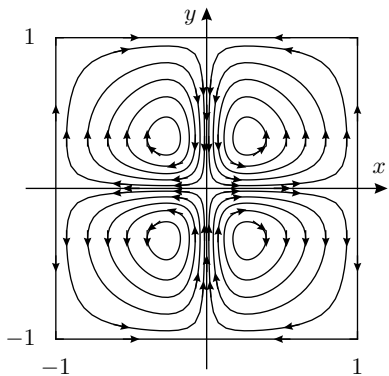}
\caption{The streaming pattern given by eq.~\eqref{Psi} is shown in one unit
  cell. 
  The streamlines circulate 
  within adjacent quarters in opposite directions. 
}
\label{f2}
\end{figure}
The positions of the atoms can then be recovered as
\be
\bx_i = \bX + \br_i(\f)= \bX + \mathbb{R}(\f) \br_i^{(0)}
\ee{r}
where the vector $\br_i(\f)$ pointing from $\bX$ 
to the
$i$th atom results from a rotation  $\mathbb{R}(\f)$ of a 
reference configuration $\br_i^{(0)}$.
\begin{figure*}[!t]
\includegraphics[height=7cm]{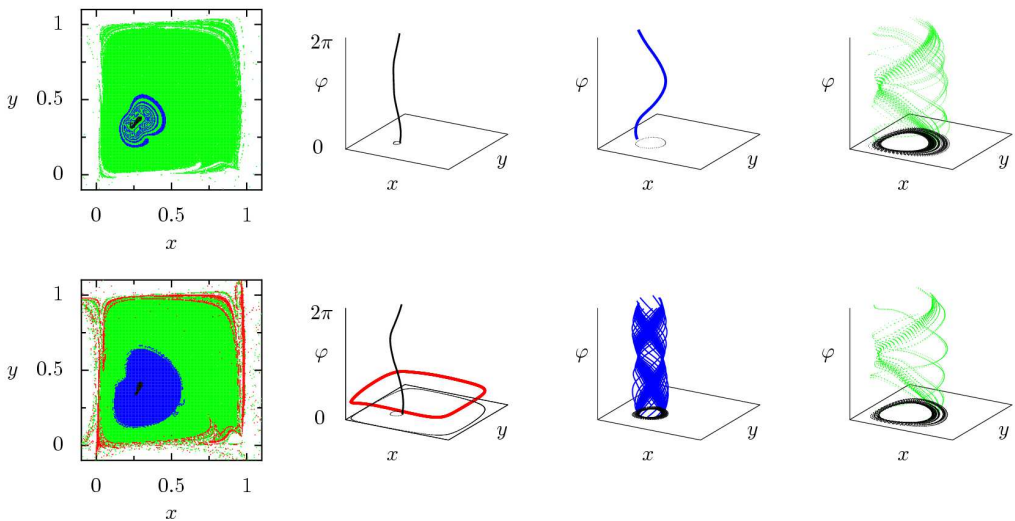}
\caption{(color). The deterministic motion of chiral partner molecules (top
vs. bottom row) is visualized by the respective domains of attraction (left
panels) and the corresponding attractors. The molecule has the shape
of an equilateral triangle with side-length d=0.2 and friction
coefficients $(\g_1,\g_2,\g_3)=(1/6,1/3,1/2)$. Its chiral partner
has the sequence $(\g_1,\g_2,\g_3)=(1/3,1/6,1/2)$. Their motions are
determined by the deterministic parts of the 
eqs.~(\ref{Xf}, \ref{Xfx}) with the velocity field
(\ref{Psi}). The left column shows cuts through the domains of
attraction at $\f=2$ within the upper right quarter of the unit cell,
see Fig.~\ref{f2}. The respective attractors are depicted in the right
panels with corresponding 
colors. Trajectories starting from white regions do not approach the
indicated quarter. Note that the domain of attraction of the
additional red attractor in the bottom row extends beyond the upper  
right quarter and consequently attracts trajectories starting outside 
of this quarter.
}
\label{f3}
\end{figure*}
The center of friction and the orientation then obey coupled Langevin
equations reading
\ba \label{Xf}
\dot{\bX} & = &\sum_i \tilde{\g}_i \: \bv(\bX+\br_i(\f)) +
\bxi_\bX,  \\
\dot{\f} & = &\sum_i \tilde{\g}_i r_{\g}^{-2} \:\br'_i(\f) \cdot \bv(\bX +
\br_i(\f)) + \x_\f
\ea{Xfx}  
The dot and the prime denote derivatives with respect to time
and angle, respectively. 
The length  
$r_{\g} = \big (\sum \tilde{\g}_i\br^2_i(\f) \big )^{1/2}$ 
is independent of $\f$ and represents 
an invariant property of the 
molecule. The fluctuating forces $\bxi_\bX(t)$ and $\x_\f(t)$ are
linear combinations of the original ones.
They vanish on average, are independent of each other, Gaussian
distributed, and therefore
characterized by their correlation functions reading
\ba
\langle \x_{\bX \a}(t) \x_{\bX \b}(s) \rangle &=& 2 
D\: \d_{\a,\b} \:\d(t-s), \\
\langle \x_\f(t) \x_\f(s) \rangle &=& 2 D \:r_{\g}^{-2} \:\d(t-s)
\ea{xXxXxf}
where the 
noise strength is given by $D= k_B T/ \sum_i \g_i$.

In distinct contrast to the two-dimensional 
velocity field $\bv (\bx)$, 
the three-dimensional vector field 
that governs the motion 
of an {\it extended} particle   
generally has a non-vanishing
divergence. It is therefore possible that in the absence of the random
forces the long-time dynamics is
ruled by one or more attractors which are approached from different
initial conditions. The actual structure of the deterministic
dynamics will depend on the properties of the
molecule and on the considered velocity
field $\bv(\bx)$. 

{\it The velocity field.}
The velocity field of an incompressible two-dimensional fluid can
always be expressed in terms of a scalar stream function $\Psi(x,y)$. 
For the sake of definiteness we here choose a periodic stream function
\be
\Ps(x,y) =  \frac{V_0 L_0  \sin(\p x/L_0) 
\sin(\p y/L_0)}{\big ( 2-\cos(\p x/L_0)
  \big )\big (3-2
  \cos (\p y/L_0))}.
\ee{Psi}
The velocity field follows as 
$v_x(\bx)  =  \partial \Ps (\bx)/ \partial y$, $ v_y(\bx) = 
- \partial \Ps (\bx)/ \partial x $.  
It is the divergence free 
solution of the Stokes equation \cite{Dhont} for a fluid that is 
driven by a quadrupolar force density
which we do not specify here. Within a unit cell ($x,y \in [-1,1]$) 
it reproduces the properties of
the experimental streaming pattern in Ref. \cite{Guttenberg04}. 
We use dimensionless variables with $L_0=1$ and $V_0=\sqrt{15}$. 
The advective transport follows the
lines of constant $\Ps$. Therefore, a unit cell contains four invariant
quarters, each containing an eddy. Because the stream function
transforms oddly under reflections at the coordinate axes, eddies in
two adjacent quarters have an opposite parity 
according to the
sense of rotation.

{\it Deterministic transport of finite particles.}
As discussed above, the motion of finite, but still small, particles 
is qualitatively
different from the advective motion of point particles.
The deterministic dynamics in the three-dimensional state space 
spanned by the two translational and
one orientational degrees of freedom is no longer conservative. It 
can be partitioned into different domains of attraction.
The detailed structure of the attractors and of their corresponding
domains of attraction depends on the size of the molecule, 
on the magnitude of the friction coefficients and also on their
sequence and therefore on the chirality of the molecule. 
for a pair of enantiomers Fig.~\ref{f3} depicts cuts of the
domains of attraction at a fixed
orientation $\f$  and the
corresponding attractors which are confined to one quarter. 
In the particular case shown in
Fig.~\ref{f3}, we found three such attractors for one enantiomer, two
of which are period-one attractors, i.e. after one full rotation of $\f$ the
molecule's center of friction has returned to its initial position.
The third attractor is chaotic. In contrast, the respective chiral
partner 
possesses four attractors in this quarter. 
Three of them are similar
to the attractors of the first enantiomer: The chaotic attractor and
one of the period-one 
attractors still exist; 
the other period-one attractor is
replaced by one  with winding number 42/43 (42 full rotations of $\f$
correspond to 43 revolutions of $\bX$). 
The fourth attractor is
periodic: Here, the orientation $\f$ performs a libration  
and the translational
degrees of freedom move in relatively close distance 
from the boundary of the considered quarter. 
This and the chaotic attractors ``collect'' also points
from adjacent quarters, see the left panel of the 
bottom row in Fig.~\ref{f3}.   
In contrast, for the chiral partner only a very small set of points
from outside the considered quarter is attracted, see the left panel
of the top row in Fig.~\ref{f3}.
For smaller molecules also more complicated
trajectories exist that stay close to the boundaries of several
quarters. However, also these trajectories differ for opposite
chiral partners and populate preferentially quarters with a parity
that is specific for the chirality of the molecule.

\begin{figure}[!t]
\includegraphics[clip=]{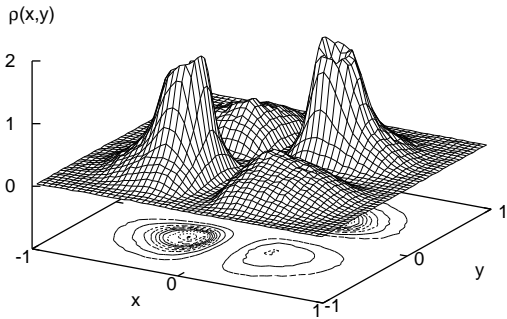}

\caption{For a single chiral molecule with size $d=0.2$
  the stationary probability 
  density integrated over all
  orientations $\f$ exhibits a pronounced
  asymmetry with respect to quarters of different parity. The friction
  coefficients are $(\g_1,\g_2,\g_3)=(1/6,1/3,1/2)$ and the dimensionless 
  noise strength
  is $D/ L_0 V_0  = 10^{-3}$.  
}
\label{f5}
\end{figure}

{\it Influence of noise.}
According to their different positions and geometric structures, the
attractors of unlike enantiomers have  different strengths 
within a quarter of definite
parity, and
consequently differ in stability with respect to thermal noise. 
At sufficiently weak noise almost all molecules settle into
the attractor with highest stability, provided the system is given
sufficient 
time to relax into its stationary state. 
Fig.~\ref{f5} depicts the resulting unequal distributions of a specific
chiral  
molecule 
in
quarters of different parity. 
For the
transport of a molecule from a quarter with the ``wrong'' to
one with the ``fitting'' parity  the attractors close to the quarters'
boundaries, like the red one in Fig.~\ref{f3}, may act as a
turnstile.      

\begin{figure}[!t]
\includegraphics[height=5cm,clip]{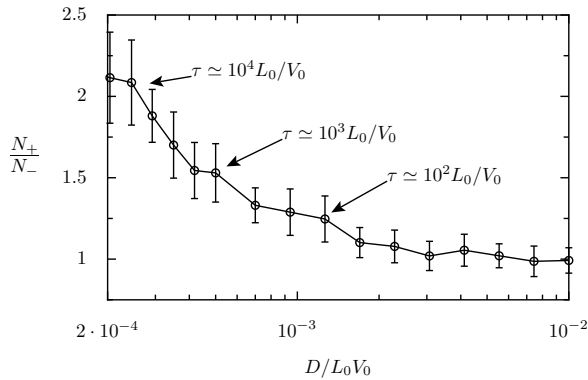}
\caption{The selectivity as given by the 
  fraction of numbers of molecules with different chirality
  $N_+/N_-$ in one quarter decreases with increasing   noise
  strength $D/L_0 V_0$. 
  The time $\tau$ it needs to reach the
  stationary value 
increases with decreasing noise
  strength. For the parameters of the molecule see
  Fig.~\ref{f3}. The error bars indicate the standard deviation as it results
  from the numerical simulation.
}
\label{f6}
\end{figure}

Because transitions from any less stable to the most stable
attractor will become increasingly rare with decreasing noise,
the approach to stationarity may proceed slowly.
On
the other hand, with higher noise levels ``wrong'' attractors will be
populated with increasing probability and therefore the separation
quality deteriorates. In Fig.~\ref{f6} the time to reach the
stationary state and the separation performance are compared as they
vary with the noise strength, which is quantified 
by the dimensionless quantity $D/ L_0 V_0 = k_B T/ \sum_i \g_i L_0 V_0$. 
It is therefore not only determined by the temperature but also by the
total friction, as well as by the characteristic eddy size and the
velocity scale. 
For example, to achieve a selectivity of $N_+/N_- =1.5$ within $15\;min$
in water at room temperature (viscosity $\h = 1g/cm\: s$) for this model flow  
an eddy size of $L_0 \approx 2\: \m m$ is required. The size of the 
molecules then is fixed to $d \approx 0.4 \m m$. This 
eddy size is smaller than what can presently be
attained 
experimentally but it should be realizable in the
near future \cite{Wixforth}.
For this rough estimate we have assumed the validity of the Stokes
law. With larger velocity gradients we also expect to improve
the separability for smaller objects.       

{\it Conclusions.}
We elucidated the transport properties of a chiral extended object in a
two-dimensional model flow in view of efficient enantiomer separation. 
As a result, we propose to start from a uniform distribution of the 
racemic mixture and wait
until the combined action of the advective transport 
and of the thermal fluctuations has led to a stationary
distribution. In this stationary state, quarters with different parity
will contain different amounts of a specific entantiomer.

There are several aspects already in two dimensions
which we have not addressed with this investigation. 
It is nevertheless clear that this model does
capture the relevant aspects of symmetry 
and of the stochastic dynamics of small but extended chiral particles. 
We are sure that 
these very aspects do also lead to chiral
separation in more realistic models describing  e.g. 
spatially extended atoms, hydrodynamic
interactions between them, as well as flexible bonds.   
Moreover, we are convinced that the same mechanisms also work in
a three dimensional helical flow again leading to enantiomer separation. 
Surface acoustic waves provide a promising tool to produce such flow patterns.
\begin{acknowledgments}
Financial support by the Deutsche Forschungsgemeinschaft via the grants
1517/13, 1517/25  and SFB 486 B13 is acknowledged.
\end{acknowledgments}

\end{document}